\documentclass{article}

\usepackage[final]{neurips_2025}

\usepackage[utf8]{inputenc} 
\usepackage[T1]{fontenc}    
\usepackage{hyperref}       
\usepackage{url}            
\usepackage{booktabs}       
\usepackage{amsfonts}       
\usepackage{nicefrac}       
\usepackage{microtype}      
\usepackage{xcolor}         
\usepackage{graphicx}
\usepackage{natbib}
\usepackage{algorithm,algorithmic}
\usepackage{wrapfig,subfigure}
\usepackage{caption}
\usepackage{multirow}
\usepackage{amsmath,amscd,amsbsy,amssymb,latexsym,bm,amsthm}
\usepackage{framed} 
\usepackage{enumitem}
\usepackage{pifont}

\usepackage{float}

\title{UniSite: The First Cross-Structure Dataset and Learning Framework for End-to-End Ligand Binding Site Detection}

\author{%
Jigang Fan$^{1,*,\dag}$ \quad
Quanlin Wu$^{1,*}$ \quad
Shengjie Luo$^{2}$ \quad
Liwei Wang$^{1,2,3,\dag}$\\
\\{$^1$}Center for Data Science, Peking University \\
{$^2$}State Key Laboratory of General Artificial Intelligence, Peking University \\
{$^3$}Center for Machine Learning Research, Peking University\\
{%
\tt\small \{jigangfan,luosj\}@stu.pku.edu.cn,\quad
\tt\small \{quanlin,wanglw\}@pku.edu.cn
}
}

\begin{document}

\maketitle
\renewcommand{\thefootnote}{\fnsymbol{footnote}}
\footnotetext[1]{Equal contribution.}
\footnotetext[2]{Corresponding authors.}

\begin{abstract}
The detection of ligand binding sites for proteins is a fundamental step in Structure-Based Drug Design. Despite notable advances in recent years, existing methods, datasets, and evaluation metrics are confronted with several key challenges: (1) current datasets and methods are centered on individual protein–ligand complexes and neglect that diverse binding sites may exist across multiple complexes of the same protein, introducing significant statistical bias; (2) ligand binding site detection is typically modeled as a discontinuous workflow, employing binary segmentation and subsequent clustering algorithms; (3) traditional evaluation metrics do not adequately reflect the actual performance of different binding site prediction methods. To address these issues, we first introduce UniSite-DS, the first UniProt (Unique Protein)-centric ligand binding site dataset, which contains 4.81 times more multi-site data and 2.08 times more overall data compared to the previously most widely used datasets. We then propose UniSite, the first end-to-end ligand binding site detection framework supervised by set prediction loss with bijective matching. In addition, we introduce Average Precision based on Intersection over Union (IoU) as a more accurate evaluation metric for ligand binding site prediction. Extensive experiments on UniSite-DS and several representative benchmark datasets demonstrate that IoU-based Average Precision provides a more accurate reflection of prediction quality, and that UniSite outperforms current state-of-the-art methods in ligand binding site detection. The dataset and codes will be made publicly available at \url{https://github.com/quanlin-wu/unisite}.
\end{abstract}

\section{Introduction} \label{Section:Introduction}
The detection of ligand binding sites on target proteins is one of the most critical steps in modern drug discovery strategies~\cite{nakashima2011structures,vincent2022phenotypic,chan2019new}. Structure-based drug design approaches begin with the three-dimensional structure of the target protein, from which deep, druggable cavities are identified. These regions, referred to as binding sites or binding pockets, are composed of sets of protein residues. Once the protein's sites are recognized, virtual screening of a molecular library can be performed using methods such as protein–ligand docking and protein–ligand affinity prediction~\cite{campbell2003ligand,li2023pocketanchor}. Alternatively, \textit{de novo} molecular design~\cite{powers2023geometric,schneuing2024structure} can be conducted based on the local structure of the binding sites to identify potential candidate compounds. As a fundamental step, the accurate identification of protein binding sites can significantly facilitate and influence subsequent steps in drug discovery. 

Over the past several decades, some endeavours have been made to detect protein ligand binding sites. These methods have evolved from traditional techniques based on geometry~\cite{fpocket2009}, template searching~\cite{findsite2008}, and energy probes~\cite{ngan2012ftsite}, to machine learning methods based on surface features~\cite{p2rank2018}, and further to deep learning methods utilizing convolutional neural networks (CNNs)~\cite{deeppocket2022} and graph neural networks (GNNs)~\cite{equipocket2023,grasp2024,vnegnn2024}. Concurrently, a series of protein–ligand datasets have also been established progressively, including scPDB~\cite{scpdb2015} and PDBbind~\cite{pdbbind2004} datasets for protein–ligand complex structures, as well as benchmark datasets such as HOLO4K~\cite{schmidtke2010large} and COACH420~\cite{yang2013protein} for evaluating binding site detection methods.

\begin{figure}[H]
  \centering
  \includegraphics[width=5.2in]{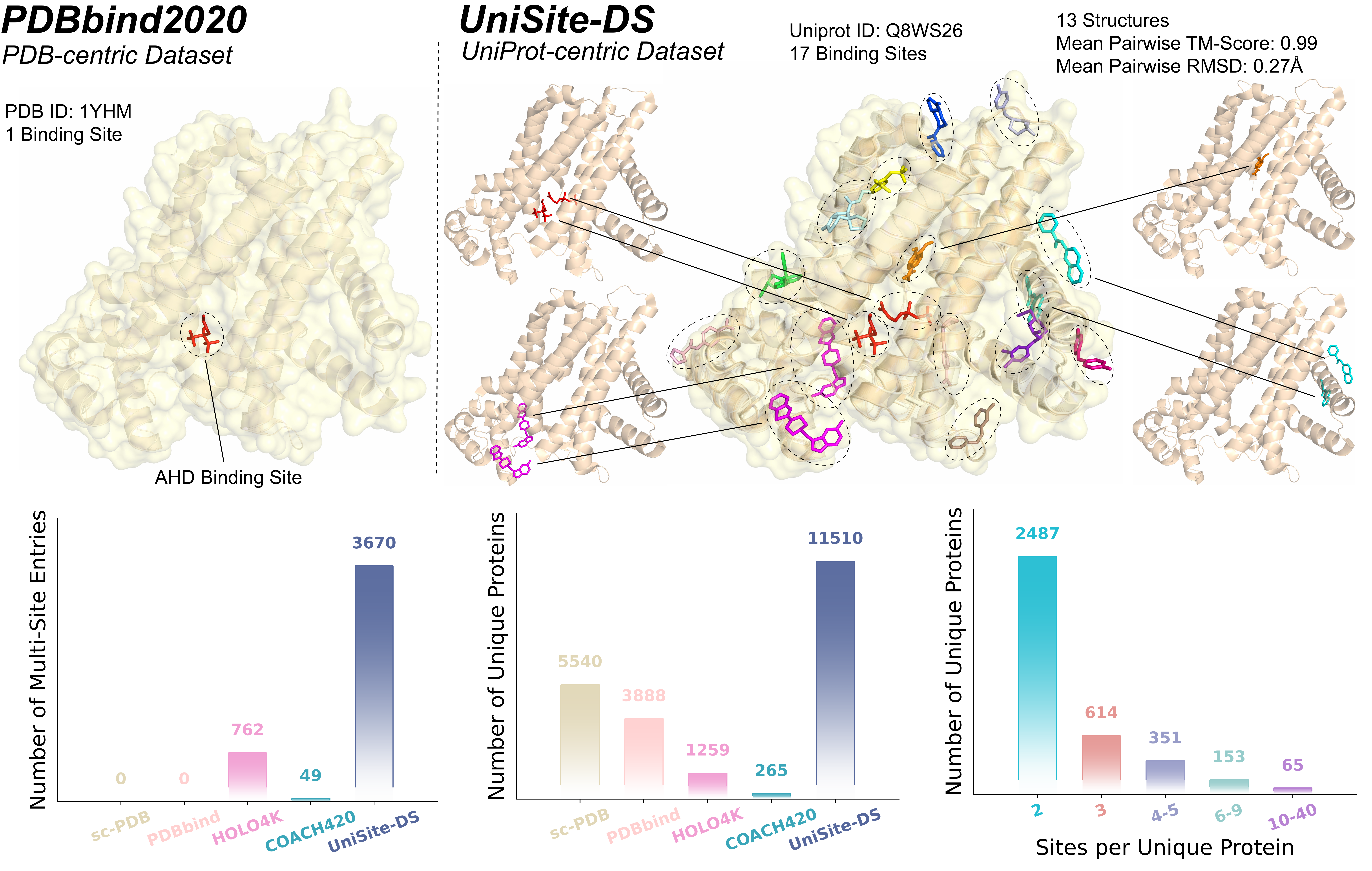}
  \caption{\textbf{Comparison between UniSite-DS and previous datasets.} \textbf{(Top left)} In PDBbind2020, only one ligand binding site and one structure are recorded for UniProt ID Q8WS26. \textbf{(Top right)} In contrast, UniSite-DS integrates distinct binding sites across all available structures (highly similar, mean TM-Score=0.99), identifying 17 unique ligand binding sites derived from 13 representative PDB entries. \textbf{(Bottom left and center)} Comparison of UniSite-DS with other widely used datasets in terms of multi-site entries and the number of unique proteins. For HOLO4K and COACH420, the most widely used \textit{mlig} subsets were selected, where each entry corresponds to a PDB structure, while in UniSite-DS, each entry corresponds to a UniProt ID. \textbf{(Bottom right)} Distribution of the number of unique proteins in UniSite-DS with respect to the number of distinct binding sites they contain.}
  \label{fig:dataset-1}
\end{figure}

Although the above efforts have significantly advanced the field of ligand binding site detection, current methods, datasets, and evaluation metrics are confronted with substantial challenges:

\textbf{Issue 1.} \textbf{All previous methods and datasets are PDB (Protein Data Bank file)-centric, specifically focusing on individual protein–ligand structures, which introduces considerable statistical bias.} Due to experimental constraints, only a limited number of binding sites in the protein are typically observed in one single protein–ligand structure where ligands are bound. However, one protein can be associated with numerous distinct protein–ligand structures, which exhibit high structural similarity in their protein components yet considerable variation in their binding site regions~\cite{he2024allosteric,matthees2024ca2,morstein2024targeting} (Figure~\ref{fig:dataset-1}). But existing datasets and methods only regard these structures as individual data entries, focusing on limited binding sites in single PDB structure. Training and evaluating on PDB-centric datasets introduces significant statistical bias, as the annotation paradigm of individual PDB structures overlooks many other ground truth binding sites.

\textbf{Issue 2.} \textbf{Existing methods employ discontinuous workflows for binding site detection.} Most approaches~\cite{fpocket2009,p2rank2018,equipocket2023,grasp2024} first perform semantic segmentation to generate binary masks of potential binding residues/atoms, then cluster them into discrete binding sites. Alternative implementations only predict binding site centers~\cite{vnegnn2024}, and the associated residues need to be extracted using external methods.
These fragmented pipelines highly rely on the post-processing methods (e.g. clustering algorithms), inherently limit end-to-end optimization and struggle with overlapping binding sites.

\textbf{Issue 3.} \textbf{Traditional evaluation metrics inadequately reflect the actual performance of binding site detection.} The most widely used evaluation metrics are DCC and DCA~\cite{p2rank2018}. DCC represents the distance between the predicted binding site center and the ground truth binding site center. DCA denotes the shortest distance between the predicted binding site center and any heavy atom of the ligand.
These metrics suffer from two fundamental limitations (Figure~\ref{fig:dcc_dca}): (1) they completely disregard the structural properties such as shape, size, and residue composition of binding sites, which are crucial for downstream tasks (Appendix~\ref{Appendix: downstream_task}), and (2) the absence of proper matching criteria between predictions and ground truth may lead to double-counting of predictions.

To address the issues mentioned above, this paper makes the following contributions:

1) We introduce \textbf{UniSite-DS}, a \textbf{manually curated, UniProt (Unique Protein)-centric} dataset of protein ligand binding sites. Leveraging the unique identifiers assigned to protein sequences in UniProt~\cite{uniprot2019uniprot}, we systematically integrated all ligand binding sites associated with given unique protein across multiple PDB structures. To the best of our knowledge, it is the first UniProt-centric dataset. Notably, UniSite-DS includes \textbf{4.81} times more multi-site proteins than existing datasets~\cite{scpdb2015,pdbbind2004}, and the overall size of the dataset is \textbf{2.08} times larger. The Uniprot-centric dataset corrects the statistical bias of previous PDB-centric datasets, thereby resolving Issue 1 and significantly broadening the available data.

2) We propose \textbf{UniSite-1D} and \textbf{UniSite-3D}, two \textbf{end-to-end} methods for protein ligand binding site detection. Both models utilize a transformer encoder-decoder architecture, supervised by a set prediction loss with bijective matching. UniSite 1D/3D directly predict $N$ potentially overlapping binding sites without requiring post-processing clustering steps, thus completely resolves Issue 2. The \textbf{UniSite-1D} variant operates exclusively on 1D protein sequence inputs, providing structure-free binding site detection capability. For enhanced performance, the \textbf{UniSite-3D} variant incorporates 3D structural information while maintaining the same end-to-end prediction framework.

3) To overcome the limitations inherent in traditional evaluation methods outlined in Issue 3, we introduce an \textbf{Average Precision (AP)} metric based on \textbf{Intersection over Union (IoU)} for fair and comprehensive binding site assessment. Extensive experiments have demonstrated that the IoU-based AP maintains strong concordance with method rankings under traditional metrics while overcoming their key limitations, providing a more accurate reflection of prediction quality.

4) Extensive experiments on UniSite-DS and classical datasets have demonstrated that our methods outperform the current state-of-the-art methods in protein ligand binding site detection. These results indicate that the end-to-end detection framework, which operates without the need for specialized feature engineering, is already capable of exhibiting strong performance for binding site detection.

\section{UniSite-DS: The First Uniprot-centric Dataset} \label{Section:UniSite-DS}
A key challenge in detecting protein binding sites is how to identify all potential binding sites~\cite{he2024allosteric,matthees2024ca2}. Most proteins contain an inherently conserved binding site, commonly referred to as the active site. The active site is shared among members of the same protein family, which means that molecules targeting this site will simultaneously target all other proteins within the family, which is highly likely to lead to off-target effects and side effects~\cite{zhang2024recent}. Identifying other binding sites within the protein that can be targeted is a crucial strategy. These sites are often located in regions topologically distant from the active site and can modulate the protein's function through allosteric effects~\cite{hughes2019asciminib,slosky2020beta,fan2021harnessing}.

As illustrated in Figure~\ref{fig:dataset-1}, one single protein can correspond to a large number of different ligand-bound structures. While the overall protein structure tends to be highly conserved, the ligand binding site regions vary considerably across these structures. The motivation behind constructing UniSite-DS lies in the recognition that identifying all potential binding sites of a protein requires a comprehensive examination of all its ligand-bound structures—an important consideration that has been overlooked by previous methods and datasets.

To construct UniSite-DS, we performed the following search and processing steps: (1) We utilized AHoJ~\cite{feidakis2022ahoj} to systematically search for all protein–ligand interactions in the PDB database~\cite{berman2003announcing}; (2) To ensure dataset quality, we excluded entries with a resolution greater than 2.5Å or those determined by non-crystallographic methods; (3) Following P2Rank's filtering criteria~\cite{p2rank2018}, we removed solvent molecules and ligands composed of fewer than five atoms, resulting in a total of 143,197 protein–ligand interaction entries; (4) For each interaction, binding site residues were identified within a 4.5Å radius of the ligand; (5) We discarded entries with three or fewer binding site residues to eliminate “floating” ligands; (6) Leveraging UniProt's unique protein sequence identifiers~\cite{uniprot2019uniprot}, we mapped binding site residues from all protein–ligand interactions of each UniProt entry to their corresponding sequences via SIFTS annotations~\cite{velankar2012sifts}, integrating all ligand binding sites across different PDB structures; (7) To eliminate data redundancy among ligand binding sites, we applied Non-Maximum Suppression (NMS) with an Intersection over Minimum (IoM) threshold of 0.7 and an Intersection over Union (IoU) threshold of 0.5, excluding highly overlapping sites. This process resulted in 13,464 distinct UniProt IDs, of which 4,846 contained multiple ligand-binding sites; (8) Based on the criteria from Proteina~\cite{geffner2025proteina}, we set the sequence length threshold to 800; (9) \textbf{We manually inspected all UniProt IDs with more than ten ligand binding sites, as well as those where a single protein–ligand complex structure contributed three or more binding sites.} As a result, we identified 11,510 valid UniProt IDs, including 3,670 with multiple ligand binding sites. The distribution of ligand binding sites is shown in Figure~\ref{fig:dataset-1}. More details about the UniSite-DS curation workflow and manual inspection process are provided in Appendix~\ref{appendix: ds_workflow}.

As the first UniProt-centric dataset, UniSite-DS encompasses \textbf{4.81} times more multi-site entries than previous datasets, and covers \textbf{2.96} times more UniProt entries than the widely used PDBbind dataset~\cite{pdbbind2004}, as well as \textbf{2.08} times more than sc-PDB~\cite{scpdb2015} (Figure~\ref{fig:dataset-1}). UniSite-DS eliminates the statistical biases inherent in earlier datasets and significantly expands the available data on multi-site ligand binding sites. Notably, case studies conducted using UniSite-DS (Appendix~\ref{Appendix: case_study}) highlighted the limitations of current binding site prediction methods in handling multi-site proteins. This observation motivated us to develop a novel end-to-end method for protein–ligand binding site detection.

\section{The Proposed Methodology}
\setlength{\intextsep}{0pt}
\begin{wrapfigure}{r}{0.5\textwidth}
  \includegraphics[width=0.5\textwidth]{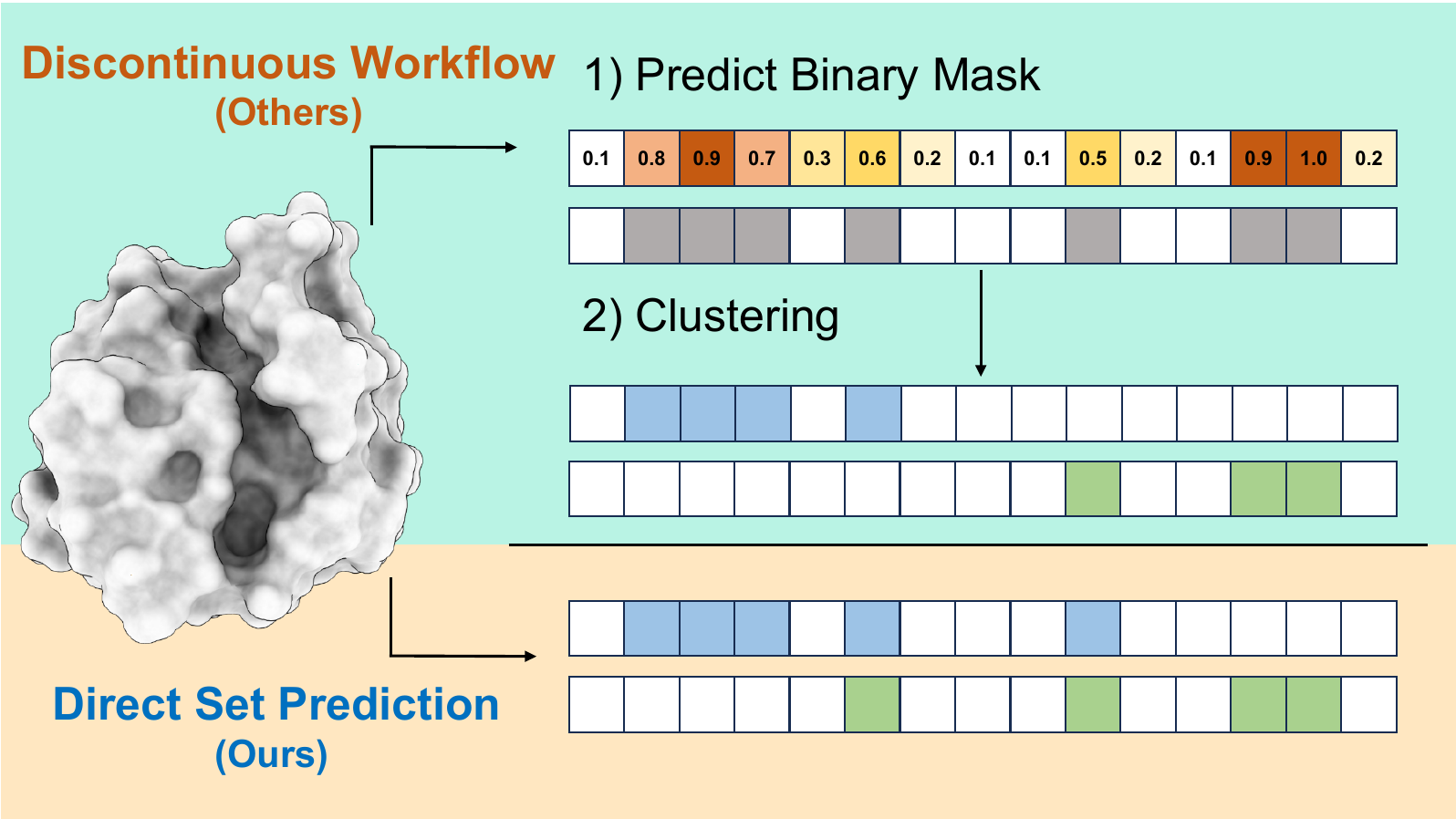}
  \caption{\textbf{Comparison of detection approaches.} \textbf{(Top)} Conventional learning-based binding site detection methods typically employ a discontinuous workflow: first predicting binary masks for residues/atoms, then clustering these masks into distinct binding sites. 
  \textbf{(Bottom)} In contrast, our method directly outputs a set of 
$N$ potentially overlapping binding sites in a single step.}
  \label{fig:pipeline}
\end{wrapfigure}

\begin{figure}
  \centering
  \includegraphics[width=\textwidth]{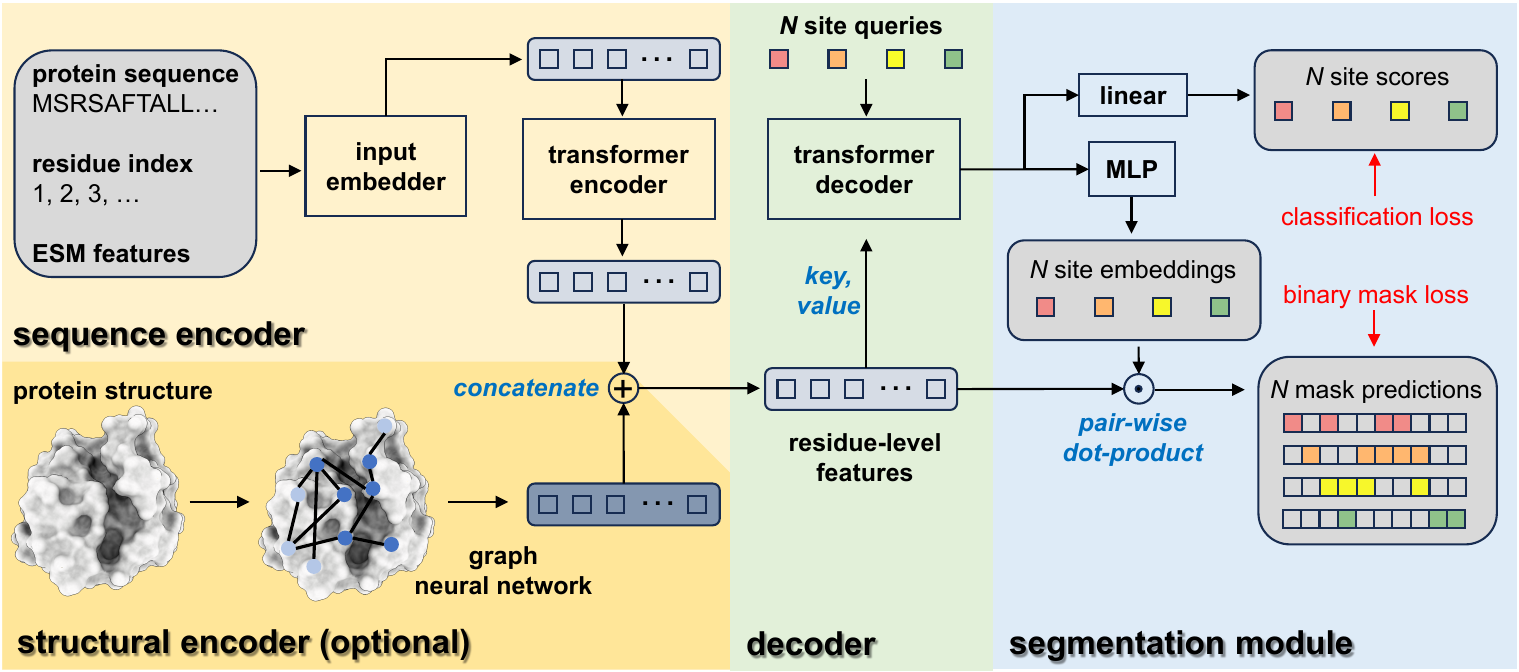}
  \caption{\textbf{The architecture of UniSite.} Our models employ an encoder to extract the residue-level features. Then a decoder module is used to generate embeddings of the $N$ predicted binding sites. Finally, the segmentation module outputs $N$ potentially overlapping binding site predictions. The encoder comprises dual pathways: a sequence encoder and an optional structural encoder, allowing UniSite to operate with either sequence-only input or combined sequence-structure information.}
  \label{fig:model arch}
  \vspace{-10pt}
\end{figure}

In this paper, we formulate protein ligand binding site detection as a \textbf{set prediction} task:
given a protein $P$ with an amino acid sequence $S$ of length $L$, the goal of binding site detection is to identify \textbf{a set of binding sites} $\{m_i^{gt}\}_{i=1}^{N_{gt}}$, where each binding site is represented by a binary mask $m_i^{gt}\in\{0,1\}^L$. Here, $m_{ij}^{gt}=1$ indicates that the $j$-th residue is part of the $i$-th site, while $m_{ij}^{gt}=0$ means it is not. Currently, most learning-based binding site detection methods adopt a discontinuous workflow: first predicting a score for each amino acid residue or heavy atom, and then clustering them into distinct binding sites. To streamline this process, we propose UniSite, the first UniProt-centric and direct set prediction approach that adheres to the end-to-end paradigm (Figure~\ref{fig:pipeline}). Two components are essential for direct set prediction in this context: (1) a set prediction loss based on bijective matching between predicted and ground truth binding sites; and (2) an architecture capable of predicting a set of sites in a single forward pass. The architecture of UniSite is shown in detail in Figure~\ref{fig:model arch}.

\subsection{Set prediction loss for binding site detection}
UniSite infers a fixed-size set of $N$ predictions $z=\{(p_i, m_i)|m_i\in\{0,1\}^L\}_{i=1}^{N}$ in a single forward pass, where $m_i$ represents the predicted binding site, and $p_i$ denotes the probability of binding and $\emptyset$ (non-binding) category. Since the ground truth set $|z^{gt}|=N^{gt}$ and the prediction set $|z|=N$ typically have unequal sizes, we assume $N\geq N^{gt}$ and pad the ground truth set with $\emptyset$ (non-binding) tokens. The padded ground truth set is defined as $z^{gt}_{pad}=\{(c_i^{gt}, m_i^{gt})|c_i^{gt}\in\{1,\emptyset\},m_i^{gt}\in\{0, 1\}^L\}_{i=1}^{N}$, where $c_i^{gt}=1$ indicates a true binding site and $c_i^{gt}=\emptyset$ corresponds to the padding. 

To train a set prediction model, we require a bijective matching $\sigma$ between the predicted set $z$ and the padded ground truth set $z^{gt}_{pad}$. 
This matching is obtained by minimizing matching cost $\mathcal{L}_{\text{match}}$:
\begin{equation}
    \hat{\sigma} = \text{arg min}\,\underset{i}{\overset{N}{\Sigma}}\, \mathcal{L}_{\text{match}}(z_i^{gt},z_{\sigma(i)})
\end{equation}
\hspace{-3pt}where $\sigma$ is a permutation of $N$ elements and $\mathcal{L}_{\text{match}}$ quantifies the pairwise matching cost between ground truth site $z_i^{gt}$ and the prediction with index $\sigma(i)$. Following prior work~\citep{detr2020, Stewart2016}, we employ the Hungarian algorithm to compute the optimal matching and use the matching cost defined as:
\begin{equation}
\mathcal{L}_{\text{match}}(z_i^{gt},z_{\sigma(i)})=-\mathbf{1}_{\{c_i^{gt}\neq\emptyset\}}\text{log}p_{\sigma(i)}(c_i^{gt})+\mathbf{1}_{\{c_i^{gt}\neq\emptyset\}}\mathcal{L}_{\text{mask}}(m_i^{gt},m_{\sigma(i)})
\end{equation}
where $\mathcal{L}_{\text{mask}}=\lambda_{\text{bce}}\mathcal{L}_{\text{bce}} + \lambda_{\text{dice}}\mathcal{L}_{\text{dice}}$ is a combination of BCE loss and dice loss~\citep{diceloss2016}, $p_{\sigma(i)}$ and $m_{\sigma(i)}$ denote the predicted probability and the binding site for the $\sigma(i)$-th prediction, respectively. This matching cost considers both the class prediction and the similarity of the predicted and ground truth binding sites. Given the optimal matching $\hat{\sigma}$, we compose a cross-entropy classification loss and the binary mask loss $\mathcal{L}_{\text{mask}}$ for each predicted site to train model parameters:
\begin{equation}
\mathcal{L}_{\text{mask\&cls}}(z^{gt},z)=\lambda_\text{cls}\sum_i^N{-\log p_{\hat{\sigma}(i)}(c_{i}^{gt})}+\mathbf{1}_{\{c_i^{gt}\neq\emptyset\}}\mathcal{L}_{\text{mask}}(m_i^{gt},m_{\hat{\sigma}(i)})
\end{equation}

\subsection{UniSite architecture}
As illustrated in Figure~\ref{fig:model arch}, UniSite comprises three main components: (1) an encoder module that extracts residue-level representations $\mathcal{F}\in\mathbb{R}^{L\times d_{\text{model}}}$ of the protein; (2) a decoder module consisting of multiple Transformer decoder layers to generate embeddings of the $N$ predicted binding sites, and (3) a segmentation module which combines the residue-level representations and the decoder embeddings to produce the final predictions $\{(p_i, m_i)|m_i\in\{0,1\}^L\}_{i=1}^{N}$. This architecture maintains conciseness while demonstrating strong compatibility with existing protein representation methods. In our implementation, we construct two variants: UniSite-1D using only sequence encoding, and UniSite-3D incorporating both sequence and structural encoders.
\vspace{-5pt}
\paragraph{Sequence encoder.} The amino acid sequence encodes the primary information of a protein and serves as the fundamental input for the UniProt-centric binding site detection. First, an input embedding module receives three inputs:
(1) learnable embeddings for the 21 amino acid types (20 standard amino acids plus an "unknown" category); (2) sinusoidal positional embedding~\citep{attention2017} for residue indices; (3) pre-trained ESM-2~\citep{esm2021} protein embeddings.
These three components are concatenated along the feature dimension, and subsequently processed by a 3-layer multilayer perceptron (MLP) to generate the initial per-residue features.
Then a stack of Transformer encoder layers process the combined features to capture residue-residue interactions and global sequence patterns.
\vspace{-5pt}
\paragraph{Structural encoder.} Protein structure serves as the most critical input for structure-based tasks, including ligand binding site detection. Recent advances have proposed various structural feature extraction approaches, including hand-crafted algorithms~\citep{p2rank2018}, CNN-based methods~\citep{deeppocket2022, deepsite2017} and graph neural network approaches~\citep{equipocket2023, vnegnn2024}. To demonstrate the generality and effectiveness of our approach, we utilize GearNet-Edge~\citep{gearnet2023}, a standard E(3)-invariant GNN model, without introducing any custom architecture or specialized feature engineering.

Given a protein $P$, the protein structure is represented as a residue-level relational graph $\mathcal{G}=(\mathcal{V}, \mathcal{E}, \mathcal{R})$, where $\mathcal{V}$ and $\mathcal{E}$ represent the set of nodes and edges respectively, and $\mathcal{R}$ is the set of edge types. Each node in the protein graph represents the alpha carbon of a residue, while sequential edges, radius edges and $K$-nearest neighbor edges are considered in the graph. Based on the defined protein graph, the node features are updated through the relational graph convolution layers~\citep{relationgraph2018} as follows:
\begin{equation}
u_i^{(l)}=\text{ReLU}\left( \text{BN}\left( \underset{r \in \mathcal{R}}{\Sigma} W_r \underset{j \in \mathcal{N}_r(i)}{\Sigma}h_j^{(l-1)}\right)\right),\,
h_i^{(l)}=h_i^{(l-1)}+u_i^{(l)}
\end{equation}
where $h_i^{(l)}$ represents the feature of node $i$ at the $l$-th layer, $\mathcal{N}_r(i)=\{j\in\mathcal{V}|(j,i,r)\in\mathcal{E}\}$ denotes the neighborhood of node $i$ with the edge type $r$, and $W_r$ is the convolutional kernel matrix shared within the edge type $r$. Specifically, BN represents a batch normalization layer and ReLU denotes the ReLU activation function.

Unlike conventional binding site detection methods that rely on structural input, our framework allows the structural encoder to be optionally included. When incorporated, the structural features are concatenated with sequence features and projected via a linear layer to match the decoder's channels.
\vspace{-5pt}
\paragraph{Transformer decoder.} The decoder comprises multiple Transformer decoder layers~\citep{attention2017} that simultaneously process $N$ embeddings of dimension $d_{\text{model}}$, $\mathcal{Q}\in\mathbb{R}^{N\times d_{\text{model}}}$, through multi-head self-attention and cross-attention mechanisms. Following established practices in~\citep{detr2020,maskformer2021}, these input embeddings are learnable positional embeddings which we refer to as \textit{site queries}. The attention mechanisms enable the decoder to perform global reasoning over all potential binding sites while incorporating contextual information from the residue-level protein features output by the encoder.
\vspace{-5pt}
\paragraph{Segmentation module.} We process the $N$ \textit{site queries} through a linear classifier followed by the softmax activation to generate class probabilities $\lbrace p_i=(p_i^{site},p_i^{\emptyset})|p_i^{site},p_i^{\emptyset}\in[0, 1]\rbrace_{i=1}^N$. Here, the classifier predicts an additional $\emptyset$ (non-binding) category to indicate when a \textit{query} does not correspond to any actual binding site. For mask prediction, the \textit{site queries} $\mathcal{Q}\in\mathbb{R}^{N\times d_{\text{model}}}$ are converted to $N$ mask embeddings $\mathcal{E}_{\text{mask}}\in\mathbb{R}^{N\times d_{\text{model}}}$ by a MLP. Finally, the binary mask prediction for each \textit{query} is computed via dot-production between the $i$-th mask embedding and the residue-level protein features $\mathcal{F}\in\mathbb{R}^{L\times d_{\text{model}}}$, followed by a sigmoid activation:
\begin{equation}
    m_i[j] = \text{sigmoid}\left(\mathcal{E}_{\text{mask}}[i, :]\cdot \mathcal{F}[j, :]^T\right)
\end{equation}

\section{Rethinking the Evaluation Metrics for Binding Site Detection}\label{sec:metric}
\begin{figure}
  \centering
  \includegraphics[width=5.2in]{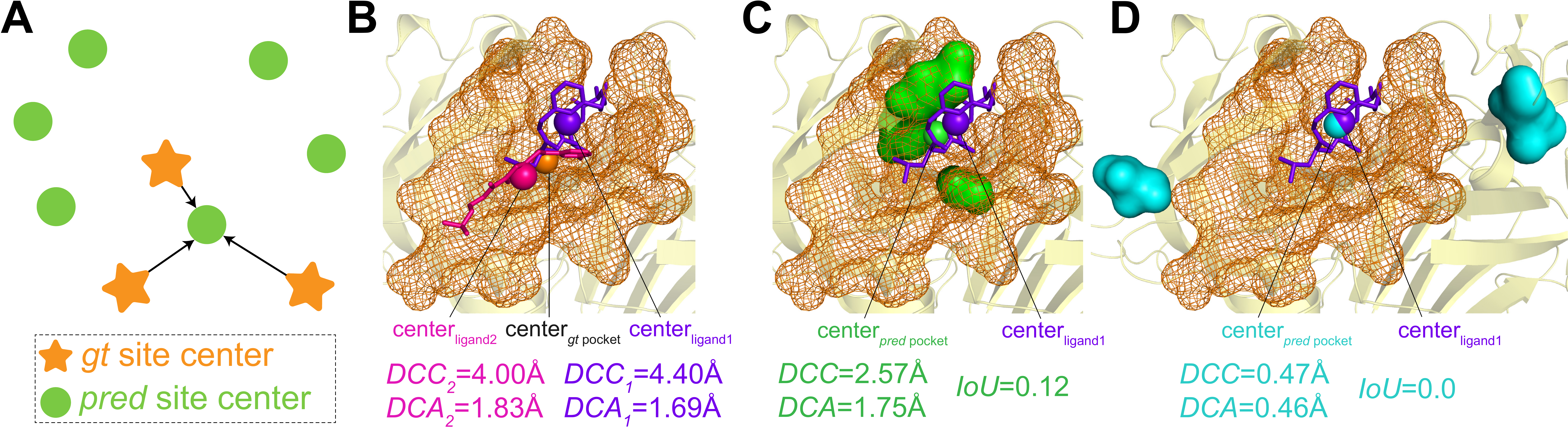}
  \caption{\textbf{DCC or DCA failure cases.} \textbf{(A)}  
  Repeated counting of the same predicted site since absence of matching. \textbf{(B)} Different ligands bound to the same site lead to deviations in DCC or DCA calculations. \textbf{(C-D)} Failed predictions classified as successful by DCC or DCA but below the IoU threshold.}
  \label{fig:dcc_dca}
  \vspace{-10pt}
\end{figure}

\textbf{DCC} (Distance between the predicted binding site center and the true binding site center) and \textbf{DCA} (Shortest distance between the predicted binding site center and any heavy atom of the ligand) are the two most widely-used metrics for binding site detection. A binding site prediction is considered \textit{successful} when its DCC or DCA value is below a predetermined threshold. Previous works~\citep{p2rank2018, deepsite2017, deepsurf2021, grasp2024}  quantify prediction performance via the \textbf{Success Rate}, defined as the ratio of \textit{successful predictions} to the total number of ground truth sites:
\begin{equation}
\text{Success Rate (DCC or DCA)} = \frac{|\{\text{Predicted sites | DCC or DCA<threshold}\}|}{|\{\text{Ground truth sites}\}|}
\end{equation}
However, these metrics suffer from two critical limitations:
\textbf{(Limitation 1)} They disregard the prediction scores or ranks, and predictions may be double-counted due to the absence of proper matching criteria (Figure~\ref{fig:dcc_dca} A and Table~\ref{tab:dc}). \textbf{(Limitation 2)} They only evaluate the center of binding sites and are ligand-dependent (DCC typically considers the ligand center as site center). Since different ligands can bound to one binding site, relying solely on ligand-centered evaluation causes these metrics to completely miss key structural properties such as the shape, size, and residue composition of the binding site. It leads to evaluation failures in certain scenarios (Figure~\ref{fig:dcc_dca} B-D), disregard the crucial information required for downstream tasks (Appendix~\ref{Appendix: downstream_task}).

The quantitative analysis of the DCC and DCA metrics is provided in Appendix~\ref{quant} to further substantiate their evaluative flaws. The analysis reveals that approximately 20\% of proteins are subject to double-counting during evaluation. Furthermore, the measured mean ground truth DCC (2.15 Å, 92.65\% < 4 Å) and DCA (1.57 Å, 98.88\% < 4 Å) exhibit a significant deviation from the ideal value of 0, revealing a systematic bias inherent to these metrics. These results directly correspond to the previously discussed limitations and confirm that DCC and DCA metrics significantly distort model performance assessment.

Previous works have recognized these limitations. To address \textbf{Limitation 1}, a common approach is to calculate the DCC or DCA for either the \textbf{top-$\boldsymbol{n}$} or \textbf{top-($\boldsymbol{n}$+2)} predicted binding sites~\citep{p2rank2018, equipocket2023, grasp2024}, where $\boldsymbol{n}$ is the number of ground truth sites. For \textbf{Limitation 2}, DeepSurf~\citep{deepsurf2021} and Utgés \textit{et al.}~\citep{cmp2024} propose computing the \textbf{IoU} between the predicted and ground truth binding site residues. Given two binding sites $m_A,m_B\in\{0,1\}^{L}$, where $L$ is the protein sequence length, the IoU of two sites is defined as:
\begin{equation}
\text{IoU}(m_A,m_B) = \frac{\text{sum}(m_A \& m_B)}{\text{sum}(m_A | m_B)}, \text{where }m_A,m_B\in\{0,1\}^{L}
\end{equation}
However, these methods fail to address the core issues, as they still lack proper matching between predicted and ground truth sites, and the \textbf{top-$\boldsymbol{n}$} or \textbf{top-($\boldsymbol{n}$+2)} metrics introduce information leakage.

To overcome these limitations, we propose to calculate the \textbf{Average Precision (AP)} metric based on the \textbf{residue-level IoU} as a fair metric for method evaluation.
We calculate AP as follows:
First, we sort all predictions by confidence scores. Then, we match each ground truth site to the predicted site with the highest score and residue-level IoU above a predetermined threshold, enforcing a one-to-one assignment constraint. Finally, we compute AP as the area under the interpolated precision-recall curve following COCO evaluation protocols~\citep{coco2014}, which is widely used in object detection. The pseudo-code for AP calculation is provided in Appendix~\ref{ap}. The AP metric offers two significant advantages: (1) the residue-level IoU enables accurate shape and size comparison between binding sites; (2) the one-to-one matching scheme inherently prevents double-counting of predictions.

\section{Experiments}
\subsection{Settings}
\vspace{-2pt}
\paragraph{Dataset.}
UniSite-DS is used for training and validation. We employ MMSeq2~\citep{mmseqs2018} to ensure that no test UniProt sequence has similarity above 0.9 to any sequence in the training set. For each UniProt sequence, we select the PDB structure with the highest sequence identity as the representative structure. Additionally, we compare UniSite with baseline methods on widely-used binding site benchmark datasets, HOLO4K~\citep{p2rank2018} and COACH420~\citep{p2rank2018}. Following DeepSurf~\citep{deepsurf2021} and EquiPocket~\citep{equipocket2023}, we use the \textit{mlig} subsets of HOLO4K and COACH420 for evaluation. Since our models are trained under a UniProt-centric schema, we only consider single-chain structures, denoting the test datasets as HOLO4K-sc and COACH420 (all structures in COACH420 are originally single-chain). All test UniProt entries are strictly excluded from the training set. More details are provided in Appendix~\ref{externel dataset}.
\vspace{-5pt}
\paragraph{Implementation details.} We set $d_{\text{model}}=256$ by default. The transformer encoder consists of 6 standard Transformer encoder layers with a feed-forward dimension 1024 and the dropout rate of 0.1. We employ 6 Transformer decoder layers following the architecture of DETR ~\citep{detr2020}. By default, We use 32 \textit{site queries}, where each query is associated with a learnable positional encoding and a zero-initialized query embedding. The multi-layer perceptron in the segmentation module consists of 2 hidden layers with 256 channels. For mask prediction, we use a combination of BCE loss and dice loss~\citep{diceloss2016}: 
\begin{equation}
\mathcal{L}_{\text{mask}}=\lambda_{\text{bce}}\mathcal{L}_{\text{bce}} + \lambda_{\text{dice}}\mathcal{L}_{\text{dice}}
\end{equation}
where $\lambda_{\text{bce}}=\lambda_{\text{dice}}=5.0$. The classification loss weight $\lambda_{\text{cls}}$ is set to $2.0$, and we downweight the classification loss by a factor of 10 when $c_i^{gt}=\emptyset$ to mitigate  class imbalance. Following DETR~\cite{detr2020}, we apply segmentation modules which share the same weights after each decoder layer, and supervise their predictions by the set prediction loss. We optimize the model using the AdamW optimizer~\citep{adamw2019} with a learning rate of $1.0\times10^{-4}$ and a weight decay factor of $0.05$. For the structural encoder, we implement the GearNet-Edge network following the origin paper~\citep{gearnet2023} without specialized feature engineering. Notably, we train the GearNet-Edge network from scratch rather than loading pre-trained weights. All models are trained on 8 NVIDIA RTX 4090 GPUs.

Our method predicts the associated residues along with a confidence score for each binding site. For applications requiring binding site centers, we compute these as the centroids of the convex hull encompassing all atoms within each predicted binding site~\cite{grasp2024, convex1970}. 
\vspace{-1pt}
\begin{table}[htbp]
  \centering
  \caption{\textbf{Results on UniSite-DS.} We highlight the top two performing methods for each metric in bold. $^{\text{a}}$ Fpocket-rescore denotes sites initially predicted by Fpocket and subsequently rescored by P2Rank. $^{\text{b}}$ VN-EGNN only outputs centers of predicted sites. For each center, We include the residues within a 9Å radius, which has the best AP performance (Appendix~\ref{vnegnn}).}
    \begin{tabular}{lccccc}
    \toprule

    \multicolumn{1}{c}{Method} & Type & Input & $\text{AP}_{0.3}$↑ & $\text{AP}_{0.5}$↑ \\
    \midrule
    Fpocket~\citep{fpocket2009} & Geometry-based & structure     & 0.1836 & 0.1017 \\
    \midrule
    Fpocket-rescore$^{\text{a}}$ & \multirow{2}[2]{*}{Machine-learning} & structure + Fpocket result   & 0.5075 & 0.2349 \\
    P2Rank~\citep{p2rank2018} &       & structure     & 0.5056 & 0.2157 \\
    \midrule
    DeepPocket~\citep{deeppocket2022} & CNN-based & structure + Fpocket result     & 0.4273 & 0.2334 \\
    \midrule
    GrASP~\citep{grasp2024} & \multirow{2}[2]{*}{GNN-based} & structure     & 0.4469 & 0.2848 \\
    VN-EGNN$^{\text{b}}$~\citep{vnegnn2024} &       & structure     & 0.1621 & 0.0705 \\
    \midrule
    UniSite-1D & \multirow{2}[2]{*}{Ours} & \textbf{sequence}     & \textbf{0.5121} & \textbf{0.3033} \\
    UniSite-3D &       & structure     & \textbf{0.5603} & \textbf{0.3835} \\
    \bottomrule
    \end{tabular}
  \label{tab:unisite-test}
  \vspace{-5pt}
\end{table}
\paragraph{Evaluation metrics.} For comprehensive evaluation, we compare IoU-based AP and traditional DCC or DCA metrics in HOLO4K-sc and COACH420. Since a Uniprot-centric data entry can contain ligands from multiply PDB structures, coordinate-dependent metrics like DCC and DCA become unsuitable due to potential inconsistencies in ligand spatial arrangements. Consequently, we merely use IoU-based AP metric on UniSite-DS. Following EquiPocket~\citep{equipocket2023}, we set the DCC or DCA threshold to 4Å, and compute the DCC or DCA success rate of \textbf{top-$\boldsymbol{n}$} predictions. We calculate AP using IoU thresholds of $0.3$ and $0.5$.

\vspace{-3pt}
\subsection{Results on UniSite-DS}\label{sec:unisite-test}
\vspace{-2pt}
The results on UniSite-DS are shown in Table~\ref{tab:unisite-test}. The geometry-based method Fpocket~\citep{fpocket2009} exhibits inferior performance since it merely considers the geometry and electronegativity. P2Rank~\citep{p2rank2018} achieves better results by extracting the protein surface features with Random Forest. DeepPocket~\citep{deeppocket2022} utilizes 3D-CNN to rescore and refine Fpocket predictions, improving the preformance in AP$_{0.5}$.
Notably, Fpocket-rescore, which combines Fpocket's initial predictions with P2Rank's re-ranking, surpasses both P2Rank and DeepPocket, highlighting the importance of proper scoring in binding site detection as captured by our AP metric. 
For graph models, GrASP~\citep{grasp2024} achieves further improvements in AP$_{0.5}$ by employing graph attention networks. However, VN-EGNN~\citep{vnegnn2024} performs poorly under AP metrics, because it only outputs predicted binding site centers, discarding structural properties (shape and size) or residue identification. For evaluation purposes, we include the residues within a 9Å radius of each predicted center, which has the best AP performance (Appendix~\ref{vnegnn}).

Trained on the UniProt-centric dataset, UniSite-1D outperforms all baseline methods without protein structure, demonstrating remarkable capability for structure-free binding site detection, particularly valuable for site-aware protein–ligand docking (Appendix~\ref{Appendix: downstream_task}). UniSite-3D further improves the performance remarkably by incorporating structure information. The above observations not only validate the effectiveness of our methods, but also reveal the significant statistical biases inherent in previous PDB-centric datasets, which limit the performance of prior methods.

\subsection{Results on HOLO4K-sc and COACH420}
The results on HOLO4K-sc and COACH420 are shown in Table~\ref{tab:benchmark}. Fpocket~\citep{fpocket2009}, P2Rank~\citep{p2rank2018} and DeePocket~\citep{deeppocket2022} exhibit a consistent performance ranking across different datasets and metrics. 
The improvement achieved by Fpocket-rescore is also consistently evident. These indicate the concordance between our proposed IoU-based AP and traditional DCC or DCA metrics. Besides, the AP metric demonstrates superior discriminative power. On HOLO4K-sc, while DeepPocket and GrASP~\citep{grasp2024} show almost identical performance in DCA$_{\text{top-}n}$ (< 0.01), they diverge substantially (> 0.10) in $\text{AP}_{0.3}$. Similarly, on COACH420, performance differences among Fpocket-rescore, P2Rank, and GrASP are more pronounced under AP evaluation than with DCC metrics.
As an exception, VN-EGNN~\citep{vnegnn2024} performs well in DCC or DCA while performing poorly under AP, as it merely predicts the binding site centers, discarding structural properties and reisidue identification. Notably, it is problematic since both the structural properties and the residue identification of binding sites are critical for downstream tasks (Appendix~\ref{Appendix: downstream_task}). 
Since UniSite-DS is a UniProt-centric dataset while HOLO4K-sc and COACH420 are both PDB-centric,
there is a training-test gap for UniSite-1D/3D. Even so, both UniSite-1D and UniSite-3D maintain strong performance on the two PDB-centric benchmarks across all evaluation metrics, demonstrating the effectiveness of our framework. More results are provided in Appendix~\ref{full benchmark}.

\begin{table}[htbp]
  \centering
  \caption{\textbf{Results on HOLO4K-sc and COACH420.} We highlight the top two performing methods for each metric in bold. $^{\text{a}}$ Fpocket-rescore denotes sites initially predicted by Fpocket and subsequently rescored by P2Rank. $^{\text{b}}$ VN-EGNN only outputs centers of predicted sites. For each center, We include the residues within a 9Å radius, which has the best AP performance (Appendix~\ref{vnegnn}).}
    \begin{tabular}{lccccccc}
    \toprule
    \multicolumn{1}{c}{\multirow{2}[4]{*}{Method}}  & \multicolumn{3}{c}{HOLO4K-sc} &       & \multicolumn{3}{c}{COACH420} \\
\cmidrule{2-4}\cmidrule{6-8}& $\text{AP}_{0.3}$↑   & $\text{DCC}_{\text{top-}n}$↑  & $\text{DCA}_{\text{top-}n}$↑  &       & $\text{AP}_{0.3}$↑   & $\text{DCC}_{\text{top-}n}$↑  & $\text{DCA}_{\text{top-}n}$↑ \\
    \midrule
    Fpocket~\citep{fpocket2009}  & 0.2711 & 0.3076 & 0.4382 &       & 0.2106 & 0.2708 & 0.4107 \\
    \midrule
    Fpocket-rescore$^{\text{a}}$  & 0.5899 & 0.5183 & 0.7654 &       & 0.5602 & 0.4405 & 0.7113 \\
    P2Rank~\citep{p2rank2018}  & 0.6011 & 0.5300  & \textbf{0.8188} &       & 0.6188 & 0.4643 & 0.7411 \\
    \midrule
    DeepPocket~\citep{deeppocket2022} & 0.5415 & 0.4925 & 0.7369 &       & 0.5184 & 0.3958 & 0.6756 \\
    \midrule
    GrASP~\citep{grasp2024}  & 0.6668 & 0.5131 & 0.7416 &       & \textbf{0.7150} & \textbf{0.4851} & \textbf{0.7620} \\
    VN-EGNN$^{\text{b}}$~\citep{vnegnn2024}       & 0.2606 & \textbf{0.5861} & 0.6999 &       & 0.2637 & \textbf{0.5446} & \textbf{0.7530} \\
    \midrule
    UniSite-1D (ours)  & \textbf{0.6867} & 0.5538 & 0.7692 &       & 0.5921 & 0.4554 & 0.7351 \\
    UniSite-3D (ours)  & \textbf{0.7091} & \textbf{0.5716} & \textbf{0.7879} &       & \textbf{0.7196} & 0.4702 & 0.7381 \\
    \bottomrule
    \end{tabular}%
  \label{tab:benchmark}%
\end{table}%
\vspace{2pt}
\begin{table}[H]
\begin{minipage}[t]{0.5\textwidth}
\makeatletter\def\@captype{table}
  \centering
  \caption{\textbf{Effect of sequence similarity.} The second column indicates the sequence identity between training sets and test sets.}
    \begin{tabular}{lccc}
    \toprule
    \multicolumn{1}{c}{Method} & Similarity & $\text{AP}_{0.3}$↑ & $\text{AP}_{0.5}$↑ \\
    \midrule
    UniSite-1D & \multirow{2}[2]{*}{<0.9} & 0.5121 & 0.3033 \\
    UniSite-3D &       & 0.5603 & 0.3835 \\
    \midrule
    UniSite-1D & \multirow{2}[2]{*}{<0.7} & 0.5056 & 0.2945 \\
    UniSite-3D &       & 0.5579  & 0.3734 \\
    \midrule
    UniSite-1D & \multirow{2}[2]{*}{<0.5} & 0.4338 & 0.2243 \\
    UniSite-3D &       & 0.4677 & 0.2801 \\
    \bottomrule
    \end{tabular}
  \label{tab:similarity}
\end{minipage}
\hspace{0.05\textwidth}
\begin{minipage}[t]{0.4\textwidth}
\makeatletter\def\@captype{table}
\centering
\caption{\textbf{Effect of site queries.} This table shows results of UniSite-3D trained with a varying number of site queries.}
    \begin{tabular}{ccc}
    \toprule
     \# of queries & $\text{AP}_{0.3}$↑ & $\text{AP}_{0.5}$↑ \\
    \midrule
    16    & 0.5515 & 0.3795 \\
    32    & 0.5603 & 0.3835 \\
    48    & 0.5562 & 0.3861 \\
    64    & 0.5615 & 0.3867 \\
    \bottomrule
    \end{tabular}
  \label{tab:query}
\end{minipage}
\end{table}

\newpage
\subsection{Ablation study}
\paragraph{Sequence similarity.}
Both the structure and function of proteins diverge with the decreasing of sequence similarity. It is necessary to evaluate our protein ligand binding site detection method across varying similarity thresholds. We employ MMSeqs2~\citep{mmseqs2018} to partition UnSite-DS with three similarity thresholds: 0.5, 0.7 and 0.9, ensuring that no test protein exceeds the corresponding similarity to any training protein. Compared to threshold 0.9, UnSite-1D/3D exhibit only a slight decrease under threshold 0.7 (Table~\ref{tab:similarity}), indicating that our methods possess generalization ability. 
As proteins with sequence similarity below 0.5 typically belong to evolutionarily distant families and exhibit markedly different structural folds, we observe significant AP performance degradation for UniSite-1D and UniSite-3D under threshold 0.5.

\paragraph{Number of site queries.} 
As shown in Table~\ref{tab:query}, UniSite-3D exhibits stable performance across varying numbers of site queries. we select 32 queries as our default configuration, considering both the computation cost and the coverage of ground truth binding sites (99.5\% proteins in UniSite-DS have sites less than 20).

\section{Concluding Remarks and Future Perspectives}

\paragraph{Key Contributions.}
In this paper, we introduce UniSite-DS, the first UniProt-centric dataset of protein ligand binding sites, which systematically integrates all ligand binding sites across multiple PDB structures for each unique protein. UniSite-DS corrects the statistical bias in previously available PDB-centric datasets and methods, while significantly broadening the available data. To amend the discontinuous workflows in existing binding site detection methods, we proposed UniSite-1D/3D, two end-to-end methods supervised by set prediction loss with bijective matching. In addition, we introduce IoU-based AP as a more accurate evaluation metric. Extensive experiments on UniSite-DS and several benchmark datasets demonstrate that our frameworks achieve superior performance, and the IoU-based AP metric can provide a more accurate reflection of binding site prediction quality.

\paragraph{Limitations and Future Work.}
The current version of UniSite-DS involves manual curation to remove unreasonable entries. A promising direction for future work is to develop automated methods for repairing and reintegrating excluded data to further enhance the dataset's coverage and quality. Additionally, our current model design aims to demonstrate the effectiveness of the end-to-end ligand binding site learning framework, without incorporating specialized feature engineering. Future investigations could explore the inclusion of specialized feature engineering to further improve model performance and generalization ability.

\section*{Acknowledgements}
We thank the reviewers for their constructive feedback, and Hannes Stärk, Chenyu Wang, Zuobai Zhang, Zaixi Zhang, and Bozitao Zhong for helpful discussions. Liwei Wang is supported by National Science and Technology Major Project (2022ZD0114902) and National Science Foundation of China (NSFC92470123, NSFC62276005).

\bibliographystyle{unsrt}
\bibliography{neurips_2025}


\newpage
\appendix
\renewcommand{\thefigure}{S\arabic{figure}}
\setcounter{figure}{0}
\renewcommand{\thetable}{S\arabic{table}}
\setcounter{table}{0}

\section{Impact of Binding Site Accuracy on Downstream Tasks}\label{Appendix: downstream_task}

\begin{figure}[H]
  \centering
  \includegraphics[width=\textwidth]{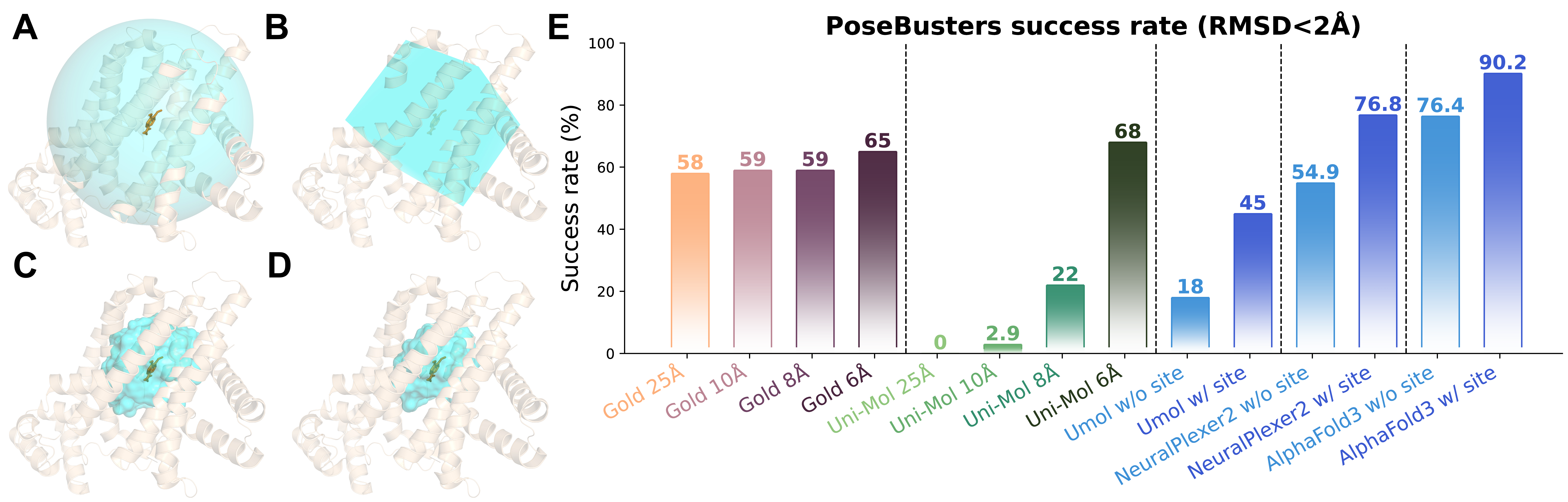}
  \caption{\textbf{The significant impact of binding site detection on molecular docking.} \textbf{(A)} Gold~\cite{verdonk2003improved} defines the binding site using a sphere. \textbf{(B)} AutoDock Vina~\cite{trott2010autodock} defines the binding site using a cube. \textbf{(C)} DeepDock~\cite{liao2019deepdock} and \textbf{(D)} Uni-Mol~\cite{zhou2023uni} identify the binding site by applying a fixed radius around the ligand. \textbf{(E)} Docking success rates on the PoseBusters dataset under different binding site configurations. Docking success rate is defined as the proportion of predictions with an RMSD less than 2Å. Data sourced from ~\cite{umol2024,buttenschoen2024posebusters,abramson2024accurate,iambic_np2}.}
  \label{fig:downstream}
\end{figure}

The accuracy of protein–ligand binding site detection is critical for downstream tasks. Taking molecular docking as an example (Figure~\ref{fig:downstream}), a comparison of different methods leads to the following conclusions:
\textbf{(1)} \textbf{The definition of binding site residues can strongly impact the docking performance}, highlighting the importance of our proposed IoU-based AP metric.
\textbf{(2)} Docking methods that do not specify binding sites (blind docking) \textbf{show significant performance improvements when provided with binding site information}, emphasizing the importance of accurate binding site identification.

\section{UniSite-DS Curation Workflow and Representative Manual Inspection Case Studies} \label{appendix: ds_workflow}

\begin{figure}[H]
  \centering
  \includegraphics[width=0.85\textwidth]{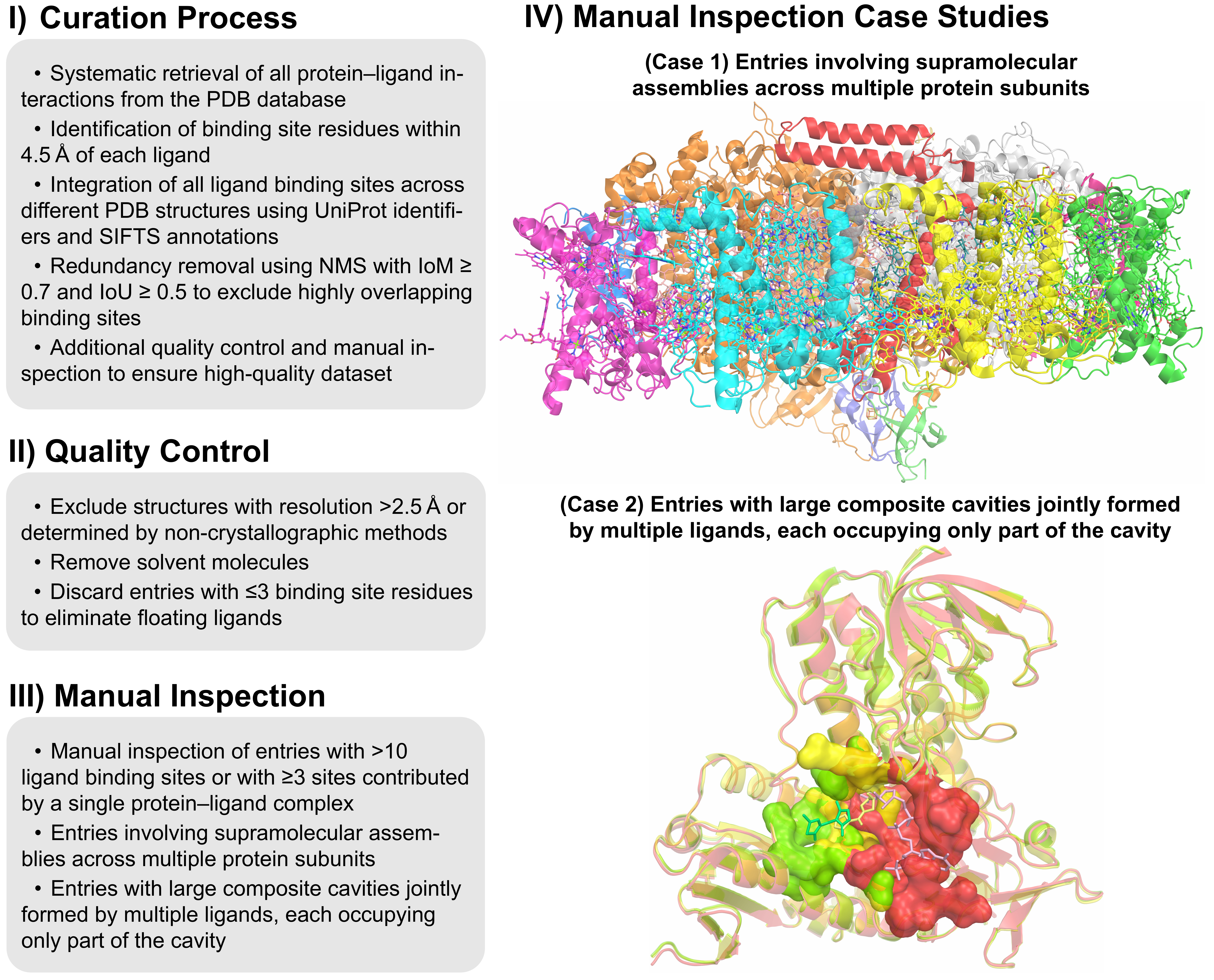}
  \caption{\textbf{Overview of the UniSite-DS workflow.} Workflow of (I) dataset curation, (II) quality control, and (III) manual inspection for UniSite-DS, together with (IV) representative manual inspection case studies.}
  \label{fig:dataset_workflow}
\end{figure}

As shown in Figure~\ref{fig:dataset_workflow}, the UniSite-DS workflow comprises three main components: (I) dataset curation, (II) quality control, and (III) manual inspection.
Below, we describe two representative types of cases encountered during manual inspection that required special consideration:

\begin{enumerate}
    \item \textbf{Case 1. Supramolecular assemblies across multiple protein subunits.}
The Photosystem I-LHCI Supercomplex is a core component of the photosynthetic machinery, consisting of multiple subunits such as UniProt ID: P05310 (Photosystem I P700 chlorophyll a apoprotein A1, PDB: 7dkz\_A), Q41038 (Chlorophyll a-b binding protein, 7dkz\_B), and Q32904 (Chlorophyll a-b binding protein 3, 7dkz\_C). Although each of these proteins binds a large number of Chlorophyll a molecules, they function as part of a tightly integrated supramolecular assembly. Therefore, they are not suitable to be included as independent entries in the UniSite-DS database and have been excluded in the current version.
    \item \textbf{Case 2. Large composite cavities jointly formed by multiple ligands.}
Trypanothione reductase (UniProt ID: Q389T8) is a key enzyme in \textit{Trypanosoma brucei} that specifically catalyzes the reduction of trypanothione, functionally analogous to glutathione reductase in mammals. The ligands from structures PDB: 5s9x\_A, 5s9t\_A, and 2wov\_C together form a large binding cavity. These ligand binding sites could potentially be merged. However, since each ligand actually occupies only a portion of the cavity rather than the entire cavity, whether such a composite cavity formed by different ligands should be considered a unified binding site often depends on system-specific definitions in the literature. As a result, entries that require further literature-based validation were excluded from the current version of the UniSite-DS database.
\end{enumerate}

\vspace{5pt}

\section{Related Work}
Over the past several decades, numerous methods have been developed for detecting protein–ligand binding sites, accompanied by advances in techniques leveraging the geometric, physical and chemical features of proteins.

Early methods relied on traditional computational algorithms. Since most binding sites show up as cavities in protein 3D structures,
geometry-based methods (Fpocket~\citep{fpocket2009}, LigSite~\citep{ligsite1997}) identify and rank these hollow cavities through hand-crafted features like alpha spheres~\citep{aphasphere1998}. Template-based methods (FINDSITE~\citep{findsite2008} and LIBRA~\citep{libra2018}) predict ligand binding sites by comparing the query protein with templates from known protein structure database. These methods typically generate a large number of predicted sites while performing poorly in ranking them. 

Subsequent approaches like PRANK~\citep{prank2015} and P2Rank~\citep{p2rank2018} employ traditional machine learning methods, particularly Random Forest. Based on the predictions of Fpocket~\citep{fpocket2009}, PRANK assigns "ligandibility" scores, which denotes ligand binding potential, to candidate sites. P2Rank is a widely used method which integrates the geometric features of the protein surface with Random Forest Algorithm.

In recent years, deep learning methods have emerged for protein–ligand binding site detection. CNN-based approaches~\cite{deeppocket2022,deepsite2017, deepsurf2021} treat protein structures as 3D images, applying 3D convolutional neural networks similar to those used in computer vision. Alternatively, GNN-based methods~\cite{vnegnn2024,grasp2024,equipocket2023} utilize graph neural networks by constructing graphs incorporating both geometric and chemical features of proteins. Despite their improved performance, these methods typically adopt discontinuous workflows: they first perform semantic segmentation to generate binary masks of potential binding residues/atoms, then cluster these masks into discrete binding sites. This fragmented pipeline heavily depends on post-processing (e.g., clustering algorithms), inherently limiting end-to-end optimization and struggling with overlapping binding sites. DETR-based architectures have also been adapted for other protein tasks, such as ProtDETR~\citep{yang2025interpretable}, which performs enzyme function classification rather than binding site detection. 

\section{Baseline Justification} \label{baseline_justification}
\vspace{-8pt}

We selected representative baseline methods that span the major methodological paradigms in ligand binding site detection. Fpocket~\citep{fpocket2009} represents traditional geometry-based computational algorithms. P2Rank~\citep{p2rank2018} is the most widely used machine learning approach. Among deep learning approaches, DeepPocket~\citep{deeppocket2022}, GrASP~\citep{grasp2024}, and VN-EGNN~\citep{vnegnn2024} were included as they represent different neural network architectures: DeepPocket employs 3D CNNs, while GrASP and VN-EGNN utilize GNNs.

For a fair and consistent comparison, we extracted ligand binding site residues from each baseline's output files according to their respective standard formats:

\begin{itemize}[leftmargin=1.5em]
    \item \textbf{Fpocket:} For each predicted pocket, Fpocket outputs a file named \texttt{pocket\{index\}\_atm.pdb}, which contains the atomic coordinates of the predicted binding site in PDB format. These atoms were directly used to identify the corresponding binding residues.
    
    \item \textbf{P2Rank:} P2Rank generates a \texttt{\{name\}.pdb\_prediction.csv} file for each input structure. The \texttt{residue\_ids} column specifies the chain IDs and residue IDs of residues constituting each predicted binding site. These identifiers were parsed to obtain the binding site residues.
    
    \item \textbf{DeepPocket:} DeepPocket refines and re-scores the pockets predicted by Fpocket, while maintaining the same output format.
    
    \item \textbf{GrASP:} GrASP outputs a \texttt{\{name\}\_probs.pdb} file, where the \texttt{bfactor} column encodes the predicted binding probabilities of heavy atoms. Following the original publication, we filtered atoms based on their predicted scores and clustered them spatially into distinct binding sites.

    \item \textbf{VN-EGNN:} VN-EGNN produces a \texttt{prediction.csv} file containing four columns (\texttt{x}, \texttt{y}, \texttt{z}, \texttt{rank}) that specify the 3D coordinates and ranking of predicted pocket centers. As the method does not directly output residue-level predictions, binding site residues were obtained by including all residues within a defined radius, as described in Appendix~\ref{vnegnn}.
\end{itemize}

\vspace{-15pt}
\section{Case Study of Classical Methods}\label{Appendix: case_study}

\begin{figure}[H]
  \centering
  \includegraphics[width=5.2in]{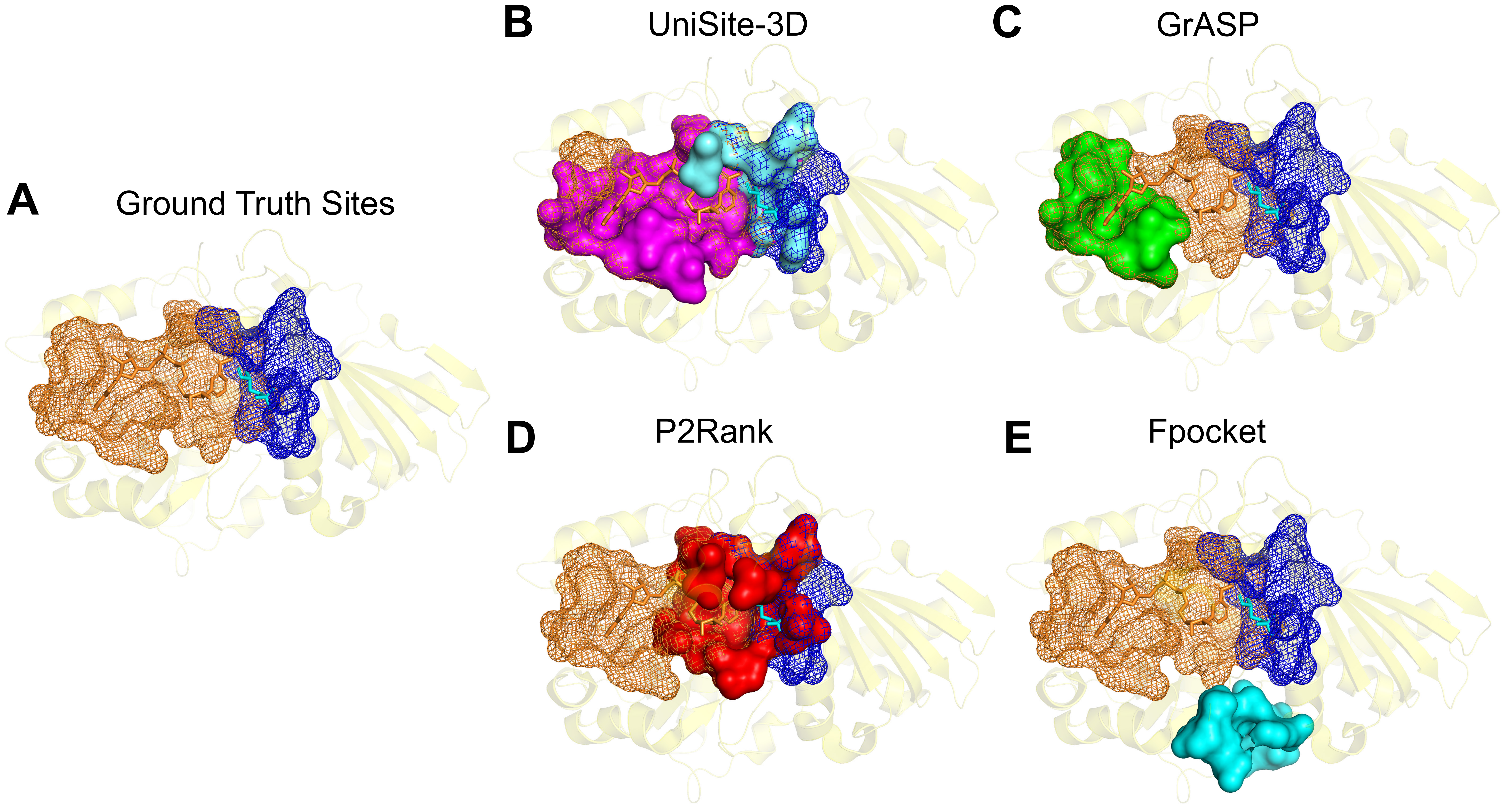}
  \caption{\textbf{Case study of classical methods.} For UniProt ID Q7YYQ9, the two ground truth binding sites are represented by dark blue and orange meshes, respectively. All predicted binding sites are shown as surfaces. \textbf{(A)} The two ground truth binding sites are colored in dark blue and orange. \textbf{(B)} The two binding sites predicted by our UniSite-3D method are colored in cyan and purple. \textbf{(C)} The single binding site predicted by the GNN-based method GrASP~\cite{grasp2024} is colored in green. \textbf{(D)} The single binding site predicted by the classical machine learning method P2Rank~\cite{p2rank2018} is colored in red. \textbf{(E)} The single binding site predicted by the geometry-based method Fpocket~\cite{fpocket2009} is colored in cyan.}
  \label{fig:case_study}
\end{figure}

\vspace{-8pt}
As discussed in Section~\ref{Section:Introduction}, \ref{Section:UniSite-DS}, previous approaches are developed based on PDB-centric datasets, which introduce substantial statistical biases. We conduct a case study of these methods, and the results (Figure~\ref{fig:case_study}) indicate that, \textbf{for multi-site proteins, existing approaches are weak in distinguishing between different ligand binding sites}. This limitation motivated us to investigate the problems in existing approaches and to develop a new methodology.

\vspace{4pt}

\section{External Datasets}\label{externel dataset}

\textbf{scPDB}~\citep{scpdb2015} is a famous dataset for protein–ligand binding site detection, commonly employed for training and validation in recent studies (~\citep{vnegnn2024,grasp2024,equipocket2023}). scPDB provides both protein and ligand structures, accompanied by the structures of binding site extracted via VolSite~\citep{volsite2018}. Notably, only one binding site and one corresponding ligand are annotated for each data entry. In this work, we use the 2017 release of scPDB, which contains 17,594 structures and 5,550 unique proteins.
(Source: \href{http://bioinfo-pharma.u-strasbg.fr/scPDB/}{http://bioinfo-pharma.u-strasbg.fr/scPDB/})

\textbf{PDBBind}~\citep{pdbbind2004} is a widely used dataset to study protein–ligand interaction, especially for protein–ligand docking~\citep{umol2024,isert2023structure}. Similar to scPDB, PDBBind annotates one ligand structure and one binding site structure in each data entry. In this paper, we use the general set of v2020, the latest academic-free edition, which comprises 19,443 structures and 3,888 unique proteins. 
(Source: \href{http://www.pdbbind.org.cn/download/}{http://www.pdbbind.org.cn/download/})

\textbf{HOLO4K} and \textbf{COACH420} are two benchmark datasets utilized for protein–ligand binding site detection. 
Follow VN-EGNN~\citep{vnegnn2024}, EquiPocket~\citep{equipocket2023} and GrASP~\citep{grasp2024}, we employ the \textit{mlig} subsets of these two dataset, which contain explicitly specified relevant ligands. HOLO4K-\textit{mlig} comprises 3,204 structures and 1,259 unique proteins, while COACH420-\textit{mlig} covers 284 structures and 265 unique proteins.
(Source: \href{https://github.com/rdk/p2rank-datasets}{https://github.com/rdk/p2rank-datasets})

\vspace{-4pt}
\section{Quantitative Analysis of Evaluation Flaws in DCC and DCA Metrics}\label{quant}
\vspace{-4pt}
In Section~\ref{sec:metric}, we discussed two critical limitations in the design of the DCC and DCA metrics. Here, we conduct a quantitative analysis on the HOLO4K-sc benchmark to demonstrate the flawed evaluation introduced by these metrics.

\textbf{Limitation 1.} The absence of proper matching criteria may lead to double-counting of predictions (Figure~\ref{fig:dcc_dca} A). We quantified the proportion of proteins affected by double-counting during evaluation on the HOLO4K-sc benchmark (Table~\ref{tab:dc}). The results reveal that DCC and DCA metrics suffer from widespread double counting artifacts, which significantly distort model performance assessment.

\begin{table}[htbp]
  \centering
  \caption{\textbf{Double counting (DC) rate of DCC or DCA metrics on HOLO4K-sc.} We highlight the top two performing methods for each metric in bold. $^{\text{a}}$ Fpocket sites rescored by P2Rank. $^{\text{b}}$ Residues within 9Å of each VN-EGNN predicted center, which has the best AP performance (Appendix~\ref{vnegnn}).}
  \label{tab:dc}
\begin{tabular}{lcc}
\toprule
\multicolumn{1}{c}{\multirow{2}{*}{Method}} & \multicolumn{2}{c}{HOLO4K-sc} \\
\cmidrule{2-3}
 & DC of $\text{DCC}_{\text{top-}n}$ & DC of $\text{DCA}_{\text{top-}n}$ \\
\midrule
Fpocket~\citep{fpocket2009} & 18.80\% & 18.31\% \\
Fpocket-rescore$^{\text{a}}$ & 13.23\% & 13.66\% \\
P2Rank~\citep{p2rank2018} & 18.98\% & 18.31\% \\
DeepPocket~\citep{deeppocket2022} & 13.53\% & 14.21\% \\
GrASP~\citep{grasp2024} & 16.29\% & 14.88\% \\
VN-EGNN$^{\text{b}}$~\citep{vnegnn2024} & 12.80\% & 11.70\% \\
\midrule
UniSite-1D (ours) & \textbf{8.88\%} & \textbf{8.94\%} \\
UniSite-3D (ours) & \textbf{9.86\%} & \textbf{10.04\%} \\
\bottomrule
\end{tabular}
\end{table}

\vspace{8pt}
\textbf{Limitation 2.} DCC or DCA only evaluate the center of binding sites and are ligand-dependent, which leads to evaluation failures in certain scenarios (Figure~\ref{fig:dcc_dca} B-D). For HOLO4K-sc, we calculated these metrics of the centroid of ground truth binding residues for each protein. The results indicate that the mean ground truth DCC is 2.15 Å (92.65\% < 4 Å), and the mean ground truth DCA is 1.57 Å (98.88\% < 4 Å). However, in principle, both DCC and DCA should ideally be 0 when evaluated using ground truth binding residues, indicating these metrics inherently contain systematic bias. 

To mitigate some of DCC's inherent limitations, we defined a corrected metric, \textbf{DCC-residue}, which uses the center of ground truth binding residues rather than the ligand center for calculation. This modification resolves failure cases caused by ligand diversity in traditional DCC evaluation. As shown in Table~\ref{tab:residue-dcc}, the corrected DCC-residue metric exhibits improved consistency with IoU-based AP in ranking the performance of different methods.

\newpage
\begin{table}[htbp]
  \centering
  \caption{\textbf{IoU-based AP, DCC-residue and DCC results on HOLO4K-sc.}
  We highlight the top two performing methods for each metric in bold. $^{\text{a}}$ Fpocket sites rescored by P2Rank. $^{\text{b}}$ Residues within 9Å of each VN-EGNN predicted center, which has the best AP performance (Appendix~\ref{vnegnn}).}
\begin{tabular}{lccc}
\toprule
\multicolumn{1}{c}{\multirow{2}{*}{Method}} & \multicolumn{3}{c}{HOLO4K-sc} \\
\cmidrule{2-4}
 & $\text{AP}_{0.3}\uparrow$ & $\text{DCC-residue}_{\text{top-}n}\uparrow$ & $\text{DCC}_{\text{top-}n}\uparrow$ \\
\midrule
Fpocket~\citep{fpocket2009} & 0.2711 & 0.2982 & 0.3076 \\
Fpocket-rescore$^{\text{a}}$ & 0.5899 & 0.5005 & 0.5183 \\
P2Rank~\citep{p2rank2018} & 0.6011 & 0.4972 & 0.5300 \\
DeepPocket~\citep{deeppocket2022} & 0.5415 & 0.4902 & 0.4925 \\
GrASP~\citep{grasp2024} & 0.6668 & 0.5379 & 0.5131 \\
VN-EGNN$^{\text{b}}$~\citep{vnegnn2024} & 0.2606 & 0.5997 & \textbf{0.5861} \\
\midrule
UniSite-1D (ours) & \textbf{0.6867} & \textbf{0.6400} & 0.5538 \\
UniSite-3D (ours) & \textbf{0.7091} & \textbf{0.6264} & \textbf{0.5716} \\
\bottomrule
\end{tabular}\label{tab:residue-dcc}
\end{table}

\vspace{-4pt}
\section{The AP Evaluation of VN-EGNN}\label{vnegnn}
\begin{figure}[H]
  \centering
  \includegraphics[width=\textwidth]{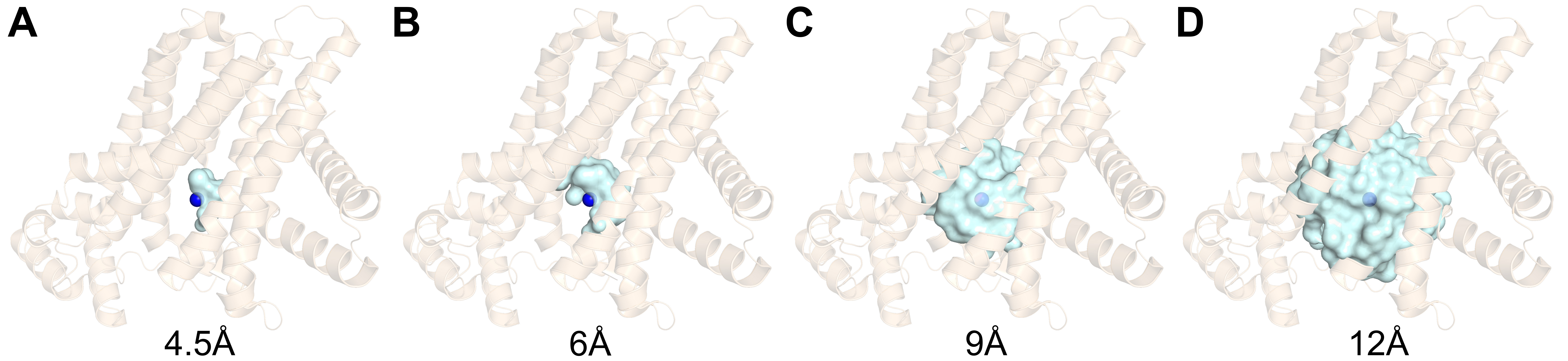}
  \caption{\textbf{Binding sites derived from the predicted center by VN-EGNN using different radii.} Binding site residues are visualized using surface representation: \textbf{(A)} residues within a 4.5Å radius; \textbf{(B)} residues within a 6Å radius; \textbf{(C)} residues within a 9Å radius; \textbf{(D)} residues within a 12Å radius.}
  \label{fig:vnegnn}
\end{figure}

\vspace{-6pt}
Since VN-EGNN~\citep{vnegnn2024} outputs only the centers of predicted binding sites, it is non-trivial to identify the binding residues. In order to evaluate the IoU-based AP, we include the residues with a fixed radius for each predicted center (Figure~\ref{fig:vnegnn}). As shown in Table~\ref{tab:vnegnn} , the radius of 9Å has the best performance across all datasets.

\begin{table}[htbp]
  \centering
  \caption{\textbf{AP evaluation results of VN-EGNN using different radii.}}
    \begin{tabular}{ccccccccc}
    \toprule
    \multirow{2}[4]{*}{Radius} & \multicolumn{2}{c}{UniSite-DS} && \multicolumn{2}{c}{HOLO4K-sc} && \multicolumn{2}{c}{COACH420} \\
\cmidrule{2-3}\cmidrule{5-6}\cmidrule{8-9}  & $\text{AP}_{0.3}$↑ & $\text{AP}_{0.5}$↑ && $\text{AP}_{0.3}$↑ & $\text{AP}_{0.5}$↑ && $\text{AP}_{0.3}$↑ & $\text{AP}_{0.5}$↑ \\
    \midrule
    4.5Å   & 0.0054 & 0.0004 && 0.0049 & 0.0001 && 0.0072 & 0.0007 \\
    6Å     & 0.0894 & 0.0088 && 0.1290 & 0.0172 && 0.1814 & 0.0189 \\
    9Å     & \textbf{0.1621} & \textbf{0.0705} && \textbf{0.2606} & \textbf{0.1346} && \textbf{0.2637} & \textbf{0.1138} \\
    12Å    & 0.0087 & 0.0010 && 0.1411 & 0.0014 && 0.1444 & 0.0013 \\
    \bottomrule
    \end{tabular}
  \label{tab:vnegnn}
\end{table}

\vspace{-5pt}
\section{The Recall Evaluation of Different Methods}
\vspace{-8pt}

The AP metric provides a comprehensive evaluation of binding site detection performance by considering both precision and recall. However, it is also important to evaluate how many true binding sites can be recovered under different IoU thresholds, since the goal of binding site detection is to identify novel and biologically meaningful sites/pockets. We computed the Recall of different methods across multiple IoU thresholds on the HOLO4K-sc dataset, and the results are summarized in Table~\ref{tab:recall}. UniSite-3D achieves state-of-the-art Recall across all IoU thresholds. 

Notably, the high recall obtained by Fpocket is due to its tendency to output nearly all potential cavities in a protein, rather than accurately identifying biologically relevant pockets. In fact, Fpocket tends to assign lower scores to true binding sites, making it challenging for biologists to distinguish meaningful pockets. This limitation has motivated many studies to develop rescoring strategies for Fpocket. While Recall reflects the potential of a method to detect biologically significant pockets, it does not account for the confidence score of each prediction. This limitation is addressed by the AP metric, which is why we adopt it as a fair and balanced criterion for evaluating different methods.

\newpage
\begin{table}[htbp]
  \centering
  \caption{\textbf{The AP and Recall results on HOLO4K-sc.} We highlight the top two performing methods for each metric in bold. $^{\text{a}}$ Fpocket sites rescored by P2Rank. $^{\text{b}}$ Residues within 9Å of each VN-EGNN predicted center, which has the best AP performance (Appendix~\ref{vnegnn}).}
\begin{tabular}{lcccccc}
\toprule
    \multicolumn{1}{c}{\multirow{2}[4]{*}{Method}} & \multicolumn{6}{c}{HOLO4K-sc} \\
    \cmidrule{2-7} & $\text{AP}_{0.3}$ & $\text{AP}_{0.5}$ & $\text{Recall}_{0.3}$ & $\text{Recall}_{0.5}$ & $\text{Recall}_{0.7}$ & $\text{Recall}_{0.9}$ \\
\midrule
Fpocket~\citep{fpocket2009} & 0.2711 & 0.1488 & \textbf{0.8361} & 0.5922 & 0.2130 & 0.0253 \\
\midrule
Fpocket-rescore$^{\text{a}}$ & 0.5899 & 0.2847 & 0.8361 & 0.5922 & 0.2130 & 0.0253 \\
 P2Rank~\citep{p2rank2018} & 0.6011 & 0.2625 & 0.7814 & 0.5337 & 0.1868 & 0.0089 \\
\midrule
DeepPocket~\citep{deeppocket2022} & 0.5415 & 0.2891 & 0.7514 & 0.5824 & 0.2584 & 0.022 \\
\midrule
 GrASP~\citep{grasp2024} & 0.6668 & 0.4126 & 0.7186 & 0.5374 & 0.2537 & 0.0159 \\
VN-EGNN$^{\text{b}}$~\citep{vnegnn2024} & 0.2606 & 0.1346 & 0.7289 & 0.4874 & 0.0566 & 0 \\
\midrule
UniSite-1D (ours) & \textbf{0.6867} & \textbf{0.4595} & 0.8212 & \textbf{0.6199} & \textbf{0.3535} & \textbf{0.0824} \\
UniSite-3D (ours) & \textbf{0.7091} & \textbf{0.5446} & \textbf{0.8469} & \textbf{0.6901} & \textbf{0.4106} & \textbf{0.1039} \\
\bottomrule
\end{tabular}\label{tab:recall}
\end{table}

\section{Full Results on HOLO4K-sc and COACH420}\label{full benchmark}
\vspace{-5pt}
\begin{table}[htbp]
  \centering
  \caption{\textbf{Full results on HOLO4K-sc.} We highlight the top two performing methods for each metric in bold. $^{\text{a}}$ Fpocket sites rescored by P2Rank. $^{\text{b}}$ Residues within 9Å of each VN-EGNN predicted center, which has the best AP performance (Appendix~\ref{vnegnn}).}
    \begin{tabular}{lcccccc}
    \toprule
    \multicolumn{1}{c}{\multirow{2}[4]{*}{Method}} & \multicolumn{6}{c}{HOLO4K-sc} \\
\cmidrule{2-7} & $\text{AP}_{0.3}$ & $\text{AP}_{0.5}$ & $\text{DCC}_{\text{top-}n}$ & $\text{DCC}_{\text{top-}n+2}$ & $\text{DCA}_{\text{top-}n}$ & $\text{DCA}_{\text{top-}n+2}$ \\
    \midrule
    Fpocket~\citep{fpocket2009} & 0.2711 & 0.1488 & 0.3076 & 0.4181 & 0.4382 & 0.5941 \\
    \midrule
    Fpocket-rescore$^{\text{a}}$ & 0.5899 & 0.2847 & 0.5183 & 0.5941 & 0.7654 & \textbf{0.8577} \\
    P2Rank~\citep{p2rank2018} & 0.6011 & 0.2625 & 0.5300  & 0.5623 & \textbf{0.8188} & \textbf{0.8652} \\
    \midrule
    DeepPocket~\citep{deeppocket2022} & 0.5415 & 0.2891 & 0.4925 & 0.5478 & 0.7369 & 0.7851 \\
    \midrule
    GrASP~\citep{grasp2024} & 0.6668 & 0.4126 & 0.5131 & 0.5267 & 0.7416 & 0.7612 \\
    VN-EGNN$^{\text{b}}$~\citep{vnegnn2024} & 0.2606 & 0.1346 & \textbf{0.5861} & 0.6339 & 0.6999 & 0.7500 \\
    \midrule
    UniSite-1D (ours) & \textbf{0.6867} & \textbf{0.4595} & 0.5538 & \textbf{0.6400} & 0.7692 & 0.8305 \\
    UniSite-3D (ours) & \textbf{0.7091} & \textbf{0.5446} & \textbf{0.5716} & \textbf{0.6470} & \textbf{0.7879} & 0.8422 \\
    \bottomrule
    \end{tabular}
\end{table}

\begin{table}[htbp]
  \centering
  \vspace{6pt}
  \caption{\textbf{Full results on COACH420.} We highlight the top two performing methods for each metric in bold. $^{\text{a}}$ Fpocket sites rescored by P2Rank. $^{\text{b}}$ Residues within 9Å of each VN-EGNN predicted center, which has the best AP performance (Appendix~\ref{vnegnn}).}
    \begin{tabular}{lcccccc}
    \toprule
    \multicolumn{1}{c}{\multirow{2}[4]{*}{Method}} & \multicolumn{6}{c}{COACH420} \\
    \cmidrule{2-7} & $\text{AP}_{0.3}$ & $\text{AP}_{0.5}$ & $\text{DCC}_{\text{top-}n}$ & $\text{DCC}_{\text{top-}n+2}$ & $\text{DCA}_{\text{top-}n}$ & $\text{DCA}_{\text{top-}n+2}$ \\
    \midrule
    Fpocket~\citep{fpocket2009} & 0.2106 & 0.1219 & 0.2708 & 0.3750 & 0.4107 & 0.5714 \\
    \midrule
    Fpocket-rescore$^{\text{a}}$ & 0.5602 & 0.2905 & 0.4405 & 0.5179 & 0.7113 & \textbf{0.8333} \\
    P2Rank~\citep{p2rank2018} & 0.6188 & 0.2618 & 0.4643 & 0.5000   & 0.7411 & 0.8034 \\
    \midrule
    DeepPocket~\citep{deeppocket2022} & 0.5184 & 0.2512 & 0.3958 & 0.4821 & 0.6756 & 0.7560 \\
    \midrule
    GrASP~\citep{grasp2024} & \textbf{0.7150} & \textbf{0.4914} & \textbf{0.4851} & 0.4970 & \textbf{0.7620} & 0.7917 \\
    VN-EGNN$^{\text{b}}$~\citep{vnegnn2024} & 0.2637 & 0.1138 & \textbf{0.5446} & \textbf{0.6071} & \textbf{0.7530} & 0.7768 \\
    \midrule
    UniSite-1D (ours) & 0.5921 & 0.2998 & 0.4554 & 0.5238 & 0.7351 & 0.8006 \\
    UniSite-3D (ours) & \textbf{0.7196} & \textbf{0.3977} & 0.4702 & \textbf{0.5387} & 0.7381 & \textbf{0.8095} \\
    \bottomrule
    \end{tabular}
\end{table}

\newpage
\section{Pseudo-code for Average Precision Calculation}\label{ap}

\begin{algorithm}
\caption{Average Precision Calculation}
\begin{algorithmic}[1]
\item[] \textbf{Input:}
\item[] \quad \texttt{prediction\_list}: A list, where each element represents the predictions for one protein. Each element is of the form $\lbrace (s_i, m_i)|s_i\in \mathbb{R} ,m_i\in\lbrace 0,1\rbrace^L\rbrace_{i=1}^{N}$, where $m_i$ represents the $i$-th predicted binding site as a binary mask of length $L$, and $s_i$ denotes the $i$-th confidence score.
\item[] \quad \texttt{ground\_truth\_list}: A list, where each element represents the ground truth (gt) binding sites for one protein. Each element is of the form $\lbrace m_i^{gt}|m_i^{gt}\in\lbrace 0,1\rbrace ^L\rbrace_{i=1}^{N_{gt}}$, where $m_i^{gt}$ represents the $i$-th ground truth binding site as a binary mask of length $L$.
\item[] \quad \texttt{iou\_threshold}: IoU threshold for considering a prediction as True Positive.

\item[] \textbf{Output:} \texttt{AP}: Average Precision under the given IoU threshold.

\quad

\textcolor{gray}{\# Step 1: Determine TP (True Positive) or FP (False Positive) for each prediction}

\quad

\FOR{\texttt{predictions\_per\_protein}, \texttt{ground\_truths\_per\_protein} \textbf{in} ZIP(\texttt{prediction\_list}, \texttt{ground\_truth\_list})}
\STATE Set all ground truths in \texttt{ground\_truths\_per\_protein} as \textbf{unused}
\STATE Sort \texttt{predictions\_per\_protein}=$\lbrace (s_i, m_i)|s_i\in \mathbb{R} ,m_i\in\lbrace 0,1\rbrace^L\rbrace_{i=1}^{N}$ by decreasing confidence score $s_i$
\FOR{$i = 1$ to $N$}
\STATE Find \texttt{ground\_truth\_j} that has the max residue-level IoU($m_j^{gt}$, $m_i$) with \texttt{prediction\_i}
\IF{IoU($m_j^{gt}$, $m_i$) $>$ \texttt{iou\_threshold} \textbf{and} \texttt{ground\_truth\_j} is \textbf{unused}}
\STATE Mark \texttt{prediction\_i} as TP
\STATE Set \texttt{ground\_truth\_j} as \textbf{used}
\ELSE
\STATE Mark \texttt{prediction\_i} as FP
\ENDIF
\ENDFOR
\ENDFOR

\quad

\textcolor{gray}{\# Step 2: Sort all predictions across all proteins by decreasing confidence scores}

\quad

\STATE \texttt{all\_predictions} $\leftarrow$ FLATTEN\_ALL(\texttt{prediction\_list})
\STATE \texttt{all\_ground\_truths} $\leftarrow$ FLATTEN\_ALL(\texttt{ground\_truth\_list})
\STATE Sort \texttt{all\_predictions} by decreasing confidence scores

\quad

\textcolor{gray}{\# Step 3: Calculate the precision-recall curve}

\quad

\STATE Set \texttt{cum\_TP}, \texttt{cum\_FP} as 0, \texttt{precision\_list} and \texttt{recall\_list} as empty
\FOR{\texttt{prediction\_i} \textbf{in} \texttt{all\_predictions}}
\IF{\texttt{prediction\_i} is marked as TP}
\STATE \texttt{cum\_TP} $\leftarrow$ \texttt{cum\_TP} + 1
\ELSE
\STATE \texttt{cum\_FP} $\leftarrow$ \texttt{cum\_FP} + 1
\ENDIF
\STATE \texttt{precision\_list}.append(\texttt{cum\_TP} / (\texttt{cum\_TP} + \texttt{cum\_FP}))
\STATE \texttt{recall\_list}.append(\texttt{cum\_TP} / LEN(\texttt{all\_ground\_truths}))
\ENDFOR

\quad

\textcolor{gray}{\# Step 4: Calculate average precision}

\quad

\STATE Calculate \texttt{AP} as the area under the precision-recall curve
\STATE \textbf{return} \texttt{AP}
\end{algorithmic}
\end{algorithm}

\end{document}